\documentclass[prl,superscriptaddress,twocolumn]{revtex4}
\usepackage[ansinew]{inputenc}
\usepackage{graphicx}
\usepackage{amsmath}
\usepackage{amsthm}
\usepackage{bm}
\usepackage{layout}
\usepackage{float}
\usepackage{txfonts}
\usepackage{amsfonts}
\usepackage{amssymb}%
\usepackage{bm}
\usepackage{graphicx}

\newcommand{\be}{\begin{eqnarray}}
\newcommand{\ee}{\end{eqnarray}}

\newcommand{\ba}{\begin{array}}
\newcommand{\ea}{\end{array}}

\begin{document}

\title{Monogamy of Bell's inequality violations in non-signaling theories}

\author{Marcin Paw{\l}owski}
\affiliation{Institute of Theoretical Physics and Astrophysics,
University of Gda{\'n}sk, PL-80-952 Gda{\'n}sk, Poland}
\affiliation{Institute of Quantum Optics and Quantum Information,
Austrian Academy of Sciences, Boltzmanngasse 3, A-1090 Vienna,
Austria}
\author{{\v C}aslav Brukner}
\affiliation{Institute of Quantum Optics and Quantum Information,
Austrian Academy of Sciences, Boltzmanngasse 3, A-1090 Vienna,
Austria} \affiliation{Faculty of Physics, University of Vienna,
Boltzmanngasse 5, A-1090 Vienna, Austria}

\begin{abstract}

We derive monogamy relations (tradeoffs) between strengths of
violations of Bell's inequalities from the non-signaling condition.
Our result applies to general Bell inequalities with an arbitrary
large number of partners, outcomes and measurement settings. The
method is simple, efficient and does not require linear programming.
The results are used to derive optimal fidelity for asymmetric
cloning in nonsignaling theories.

\end{abstract}
\maketitle

The non-signaling principle -- the impossibility of sending
information faster than the speed of light -- is deeply rooted in
our existing understanding of the physical world. It not only allows
to consider current physical theories within a general framework of
the non-signaling principle, but also to significantly restrict the
structure of possible future theories. This principle implies that
the correlations between distant partners cannot be used to send
information, as is the case for quantum correlations.
Mathematically, a correlation is defined as a joint probability
distribution $P(a, b|x, y)$, where $a$ and $b$ are outcomes of two
separated parties, say Alice and Bob, given $x$ and $y$ are their
free choices of measurement settings, respectively. The
non-signaling condition implies that the marginals are independent
of the partner's choice: $P(a|x, y) =\sum_{b} P(a, b|x, y) =
P(a|x)$.

Quantum theory predicts correlations between space-like separated
events, which are non-signaling but cannot be explained within local
realism, i.e. within the framework in which all outcomes have
pre-existing values for any possible measurement before the
measurements are made (``realism'') and where these values are
independent from any action at space-like separated regions
(``locality'')~\cite{bell}. This is signified by the violation of
Bell's inequalities. Since the work of Popescu and
Rohrlich~\cite{PS} it is known that there are correlations violating
Bell's inequality stronger than the quantum mechanical correlations,
but without contradicting the non-signaling principle. This opened
up a possibility to investigate quantum correlations outside of the
Hilbert space formalism as well as correlations in general
probabilistic theories subject to the non-signaling
constraint~\cite{BLM+,MAG,Barrett,Barnum+,BBLW}.

The general framework for considering non-signaling correlations is
also important from the information-theoretical point of view. For
example, protocols for a secret key distribution were recently
proposed and their security proved solely using the non-signaling
principle~\cite{BHK,MAG2}. Furthermore, it was shown that every
non-signaling theory that predicts the violation of Bell's
inequality implies the no-cloning theorem. The bound on the
shrinking factor for the symmetric, phase-covariant cloning was
derived from the non-signaling condition~\cite{Werner,MAG}.

In this paper we will investigate monogamy properties of
correlations in non-signaling theories. This property was first
found for quantum entanglement. Consider, for example, three
subsystems A, B and C of a composite quantum system. The theorem of
Coffman, Kundu, and Wootters describes the tradeoff between the
degree of entanglement between A and B, and the degree of
entanglement between A and C, as measured by
concurrence~\cite{CKW,OV,Koashi,Hiroshima}. A similar trade-off
exits between the violation of the Clauser-Horne-Shimony-Holt (CHSH)
inequality for the pair A-B and the violation of the inequality for
the pair A-C in {\it any} non-signaling theory~\cite{Toner,MAG} (For
the trade-off derived within quantum theory see Ref.~\cite{SG,TV}).
The questions arise: Is the monogamy relation a generic feature of
{\it every} Bell inequality? What are constraints on quantum
correlations imposed by the non-signaling condition? A general, but
only qualitative result was found~\cite{MAG}: If A and B maximally
violate some Bell inequality, then A and C are completely
uncorrelated. Furthermore, a linear programm was given for finding
the non-signaling bounds on the quantum value of a general Bell
expression~\cite{Toner}.

Here we derive the monogamy relations for the violation of {\it
general} Bell's inequalities in any non-signaling theory. It applies
for an {\it arbitrary} number of parties, measurement settings and
outcomes. The method is simple, efficient and does not require
linear programming. To illustrate its applicability we derive the
optimal fidelity for generally asymmetric cloning from the
non-signaling bounds. The latter generalizes the results of
Ref.~\cite{MAG}

Consider a general linear, two-partite Bell inequality, for
correlations of local outcomes observed at measurement stations of
Alice (A) and Bob (B): \be \label{bell} {\cal B}(\text{A},\text{B})
\equiv \sum_{x,y} \sum_{a,b} \alpha(x,y,a,b) P(A_x=a,B_y=b) \leq R.
\ee Here $x$ and $y$ stand for the measurement settings chosen by
Alice and Bob respectively, and $a$ and $b$ for the outcomes of
their measurements.
$R$ is the local realistic bound and $P(A_x=a,B_y=b) \equiv
P(a,b|x,y)$ is the conditional probability (both notations will be
used in the present work).

Throughout this Letter, we will assume that every Bell inequality is
written in such a form that for all $x,y,a,b$ one has $
\alpha(x,y,a,b)\geq 0 $. This guarantees that ${\cal
B}(\text{A},\text{B})\geq 0$ and $ R\geq 0$. To see that every
inequality can be brought in this form note that each inequality
which has some negative $\alpha$'s can be rewritten by substituting
probabilities which are next to negative $\alpha$'s by unity minus
the probability of the opposite events. The chosen form simplifies
the formulas for Bell's inequalities as no absolute values need to
be involved.

We now give the main result of our paper. Consider $n + 1$ separated
parties, a single Alice (A) and a set of $n$ Bobs
$(\text{B}^{(1)},...,\text{B}^{(n)})$. Furthermore, consider a
linear bipartite Bell's inequality ${\cal
B}(\text{A},\text{B}^{(m)}) \leq R$ of type~(\ref{bell}), for
measurements of A and any single Bob $\text{B}^{(m)}$, $m\in
\{1,...,n\}$. The number of outcomes at the two stations is
arbitrary, as well as the number of measurement settings at A. The
number of settings at each $\text{B}^{(m)}$ is assumed to be $n$,
which is also the total number of Bobs. The following monogamy
relation must hold between the strengths of violations of bipartite
Bell's inequalities for $n$ pairs of observers, each pair consisting
of Alice and single Bob: \be \label{sig} \sum_{m=1}^n {\cal
B}(\text{A},\text{B}^{(m)})\leq n R. \ee This holds in every
non-signaling theory, including these for which individual Bell's
inequalities ${\cal B}(\text{A},\text{B}^{(m)}) \leq R$ can be
violated, as it is the case in quantum theory (An analogous result
of Eq.~(\ref{sig}) within quantum theory was found in
Ref.~\cite{Barbara}.).

The proof consists in showing that a violation of the monogamy
relation~(\ref{sig}) would imply signaling. The left-hand side of
Ineq.~(\ref{sig}) can be written as $\sum_{m=1}^n{\cal
B}(\text{A},\text{B}^{(m)})=\sum_{m=1}^n {\cal B}_m $, where \be
{\cal B}_m \equiv \sum_{x,y} \sum_{a,b} \alpha(x,y,a,b)
P(A_x=a,B_y^{(y+m-1\text{mod} n)}\!=b) \ee involves a sum over all
the settings of Alice and only one setting for each Bob (see Figure
1). Here $P(A_x=a,B_y^{(y+m-1\text{mod} n)}\!=b)$ is the probability
that Alice observes $a$ and the $(y+m-1 \mod n)$-th Bob observes
$b$, when she chooses setting $x$ and he setting $y$. If
Ineq.~(\ref{sig}) is violated, then there exists at least one $m$
for which \be \label{polr} {\cal B}_m \leq R \ee is violated. We
show that violation of Ineq.~(\ref{polr}) implies signaling. We
prove it for $m=1$, for other $m$ values the proof is analogous. The
inequality~(\ref{polr}) for $m=1$ reads \be \label{polr2} {\cal B}_1
\equiv \sum_{x,y}\sum_{a,b}\alpha(x,y,a,b)P(A_x=a,B_y^{(y)}=b)\leq
R. \ee and is again a Bell's inequality of type~(\ref{bell}).

It is important to note that in the present set-up Bobs do not
change their measurement settings during the Bell test and,
furthermore, that they all can jointly perform their measurements.
Thus, observer $\text{B}^{(1)}$ always performs measurement 1, and
simultaneously $\text{B}^{(2)}$ performs measurement 2 and so on.
Let us introduce the joint probability
$P(A_x=a,B_1^{(1)}=b_1,...,B_n^{(n)}=b_n) $ that Alice observes the
outcome $a$ when she chooses the setting $x$, and Bobs observe
sequence of outcomes $b_1,...,b_n$. We write $ P(A_x=a,B_y^{(y)}=b)=
\sum'_{b_1,...,b_n}
P(A_x\!=\!a,B_1^{(1)}\!=\!b_1,..,B_y^{(y)}\!=\!b,..,
B_n^{(n)}\!=\!b_n) $, where $\sum'_{b_1,...,b_n}$  denotes the sum
over all indices $b_1,...,b_n$ except $b_y$.  The
Ineq.~(\ref{polr2}) can now be brought into the form:
\begin{eqnarray} \nonumber
{\cal B}_1 =
\lefteqn{\sum_{x,a} \sum_{b_1,...,b_n} \alpha'(x,a,b_1,..,b_n) \times}  \\
& & P(A_x\!=\!a,B_1^{(1)}\!=\!b_1,...,B_n^{(n)}\!=\!b_n) \leq R,
\label{polr3}
\end{eqnarray} where $\alpha'(x,a,b_1,..,b_n)=\sum_y \alpha(x,y,a,b_y)$.

\begin{figure}
\includegraphics[width=6.5cm]{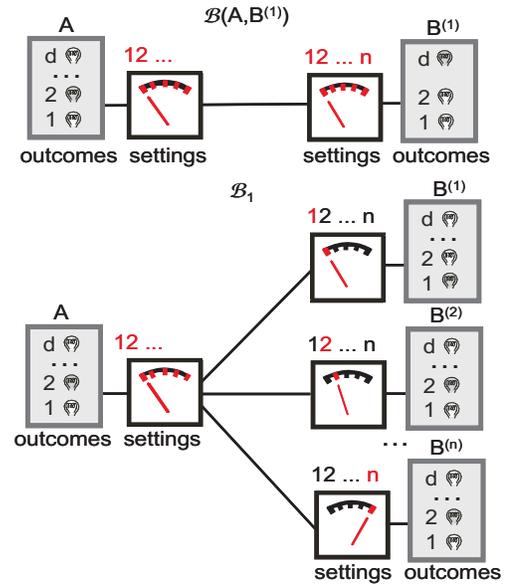}
\caption{Diagram of measurements involved in the Bell expression
${\cal B}(A,B^{(1)})$ (top) and ${\cal B}_1$ (bottom). The choices
of the measurement settings are marked in red color (online) and by
positions of the pointers (print). In the set-up for ${\cal
B}(A,B^{(1)})$ both parties have a number of measurements to
(freely~\cite{kofler}) choose from. In the set-up for ${\cal B}_1$
only Alice has such a choice whereas each $B^{(y)}$, $y\in
\{1,...,n\}$ always performs the same measurement $y$.}
\end{figure}

We introduce the short notation $\vec{b} \equiv (b_1,...,b_n)$ for
the set of all outcomes that are observed by Bobs and
$P(a,\vec{b}|x) \equiv
P(A_x\!=\!a,B_1^{(1)}\!=\!b_1,..,B_y^{(y)}\!=\!b,..,B_n^{(n)}\!=\!b_n)$
for the probability that Alice observes $a$ and Bobs $\vec{b}$
conditional on her choice of setting $x$. Recall that these
probabilities are not conditioned on the choice of the measurement
settings of Bobs since in the set-up considered (${\cal B}_1$) all
settings $y$ are chosen simultaneously by different Bobs.  We now
can rewrite Ineq.~(\ref{polr3}) as \be \label{polr4} {\cal B}_1 =
\sum_{x,a,\vec{b}} \alpha'(x,a,\vec{b}) P(a,\vec{b}|x)\leq R. \ee
For every probability distribution it is valid that \be
\label{saban} P(a,\vec{b}|x)=P(a|\vec{b},x)P(\vec{b}|x). \ee The
non-signaling condition is the assumption that \be
P(\vec{b}|x)=P(\vec{b}), \ee which allows to write Eq.~(\ref{saban})
as \be  \label{prob} P(a,\vec{b}|x)=P(a|\vec{b},x)P(\vec{b}). \ee

It is crucial to realize that a probability distribution that
satisfies Eq.~(\ref{prob}) is explainable within local realism. In a
local realistic model the source sends particles carrying
information about the vector $\vec{b}$ with the probability
$P(\vec{b})$ to Alice and all Bobs. The measurement apparatuses of
Bobs output $\vec{b}$ while the apparatus of Alice takes the input
$x$ (the setting chosen freely by Alice) and outputs $a$ with the
probability $P(a|\vec{b},x)$. This means that every value of the
left-hand side of Ineq.~(\ref{polr4}), which is attainable by any
non-signaling theory, is also attainable by a local realistic one.
And since $R$ is the maximal attainable value of the left-hand side
of Ineq.~(\ref{polr4}) this in turn means that a violation of
Ineq.~(\ref{polr4}) would allow Alice to signal to Bobs.

The extension to multipartite Bell's inequalities is
straightforward. Consider Bell's inequality \be \label{multip} {\cal
B}(\text{P}^{(1)},...,\text{P}^{(N)} )\leq R \ee where $N$ parties
$\text{P}^{(i)} $, $i \in \{1,...,N\}$, can choose between an
arbitrary number of measurement settings. We can always divide the
parties into two sets and name these sets $\vec{A}$ and $\vec{B}$.
We can now consider each of these two sets as one party in a
corresponding two-partite Bell's inequality and rewrite
Ineq.~(\ref{multip}) as ${\cal B}(\vec{A},\vec{B})\leq R$. Each
setting of $\vec{A}$ and $\vec{B}$ corresponds to one of all the
possible combinations of
 settings of individual parties that form the set. Following the
proof given above we can conclude: For every $N$-partite Bell
inequality $ {\cal B}(\text{P}^{(1)},...,\text{P}^{(N)} )\leq R $
and any chosen division of the parties into two sets $\vec{A}$ and
$\vec{B}$, the violation of \be \label{gen1} \sum_{m=1}^{n}{\cal
B}(\vec{A},\vec{B}^{(m)})\leq n R, \ee where $n$ is the number of
settings at each $\vec{B}^{(m)}$ (the number of the settings at $A$
is arbitrary), implies signaling.

To illustrate the consequences of our result we will consider an
asymmetric, state dependent cloning machine (see ref.~\cite{SIGA}
for a review on cloning) that takes a single system of arbitrary
dimension and produces $n$ copies ($1\rightarrow n$ cloning
machine). We will derive the optimal shrinking factor for the
machine from our non-signaling inequalities~(\ref{sig}). Consider a
composite system consisting of two subsystems belonging to A and B.
The two subsystems can be measured locally giving rise to the
probability distribution $P(\text{A}_x=a,\text{B}_y=b)$ (Figure 2,
top). Alternatively, the subsystem of B can be sent to the cloning
machine which takes it as an input and outputs $n$ ``copies'' that
are further distributed to $n$ observers $\text{B}^{(m)}$, $m \in
\{1,...,n\}$, and then measured locally in coincidence with the
subsystem of A. The ``cloned'' probability distribution for local
measurements on A and $\text{B}^{(m)}$ is denoted by
$P(\text{A}_x=a,\text{B}^{(m)}_y=b)$. Given an initial probability
distribution, which cloned probability distributions are in
agreement with the non-signaling condition?

\begin{figure}[!h]
\includegraphics[width=7.3cm]{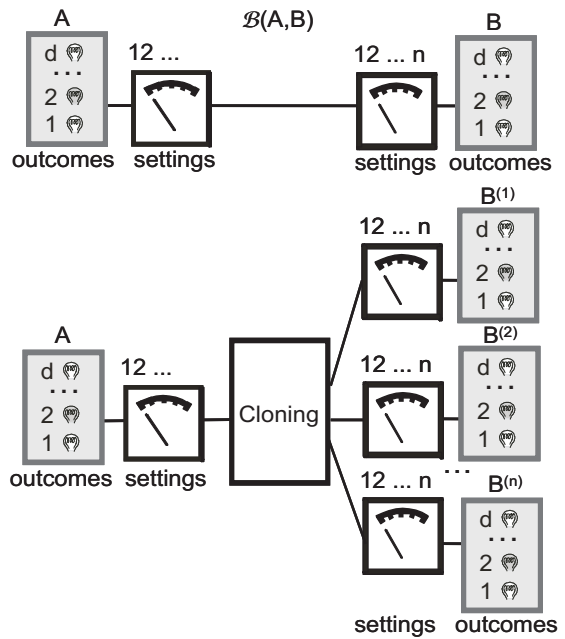}
\caption{Diagram of measurements involved in a direct Bell test
between Alice and Bob (top) or the one between Alice and $n$ Bobs
after the cloning procedure (bottom). While Alice chooses between an
arbitrary number of measurement settings, Bobs choose between $n$ of
them. The non-signaling condition gives an upper bound of the
shrinking factor for a general asymmetric cloning procedure employed
by Bobs.}
\end{figure}

We now compare the strengths of the violation of Bell's inequality
on an arbitrarily dimensional composite system before and after the
cloning procedure. Denote the Bell expression in the experiment
without cloning by ${\cal B}(\text{A},\text{B})$. Alice is assumed
to choose between an arbitrary number of measurement settings and
Bob between $n$ of them. Denote, furthermore, the Bell expressions
in the experiment with cloning by ${\cal
B}(\text{A},\text{B}^{(m)})$, $m \in \{1,...,n\}$. Every such
expression involves the cloned probability distribution between a
pair of observers $A$ and $\text{B}^{(m)}$, where $A$ chooses among
an arbitrary number of measurement settings, and each
$\text{B}^{(m)}$ between $n$ of them. We define the mean value of
the shrinking factor $\eta_m$ for each of the copies to be \be
\eta_m=\frac{{\cal B}(\text{A},\text{B}^{(m)})}{{\cal
B}(\text{A},\text{B}).} \ee The non-signaling inequality~(\ref{sig})
implies \be \sum_{m=1}^n {\cal B}(\text{A},\text{B}^{(m)}) \leq n R,
\ee which transforms into \be \frac{1}{n} \sum_{m=1}^n \eta_m \leq
\frac{R}{{\cal B}(\text{A},\text{B})}. \ee Therefore, the bound on
the mean value of the shrinking factor for cloning is non-trivial,
i.e. less than unity, only if the initial probability distribution
{\it violates} Bell's inequality. This generalizes the results of
Ref.~\cite{MAG} obtained for the symmetric cloning and $n=2$.

The bound derived with the use of the CHSH inequality ($n=2$) is
$\frac{1}{\sqrt{2}}$ and is, interestingly, shown to be saturated by
quantum mechanics~\cite{MAG}. Using our result every two-partite
Bell's inequality which provides an upper bound on the Grothendieck
constant $K_G(d)$ for $d$-dimensional systems and involves $n$
settings at one of the parties (if the numbers are different for
different parties, any number can be taken) gives a bound of
$\frac{1}{K_G(d)}$ on the shrinking factor of symmetric
$1\rightarrow n$ cloning machine. For example, the recent result
\cite{HG} provides us with stronger bounds on the shrinking factors
for the symmetric cloning of qubits for cloning machines that make a
very large number of copies (at least $1\rightarrow 465$).

In conclusion, we derive monogamy constraints on correlations using
only non-signaling condition. Our results can be applied to any
Bell's inequality and in each the case the constraints they give are
easy to calculate. These constraints hold for every non-signaling
theory, quantum mechanics being a special case. This generalizes
previously known results which either were explicitly obtained for
only the CHSH inequality or gave only qualitative description of
monogamy. We also have shown an exemplary application of our results
in finding a straightforward way to derive bounds on shrinking
factors of cloning machines in any non-signaling theory.

This work has been supported by the FWF within Projects No.
P19570-N16, SFB and CoQuS (No. W1210-N16), and EC Project QAP (No.
015846). MP has partially done this work at National Center for
Quantum Information of Gdansk.

\end{document}